\documentclass[aps,prl,twocolumn,showpacs,dvipsnames,superscriptaddress]{revtex4-2}

\usepackage[utf8]{inputenc}

\usepackage{amssymb,amsfonts,amsmath} 
\usepackage{graphicx,epsfig,psfrag}
\usepackage{color}
\usepackage{natbib}
\usepackage{url}
\usepackage[breaklinks=true]{hyperref}
\usepackage{mathtools}
\usepackage{subfigure}
\usepackage{physics}
\usepackage{lipsum}
\usepackage{comment}
\hypersetup{
        colorlinks = true,
        citecolor = blue
}

\begin{document}

\title{Quantum Beats in Many-Body Localized Systems}

\author{Bernard Faulend}
\altaffiliation{Present address: Institute for Theoretical Physics and Vienna Center for Quantum Science and Technology, Vienna University of Technology (TU Wien), 1040 Vienna, Austria}
\author{Hrvoje Buljan}
\author{Antonio \v{S}trkalj}
\email{astrkalj@phy.hr}
\affiliation{\mbox{Department of Physics, Faculty of Science, University of Zagreb, Bijeni\v{c}ka c. 32, 10000 Zagreb, Croatia}}

\begin{abstract}
Slow particle dynamics observed numerically even deep inside the many-body localized (MBL) regime has called the stability of MBL in the thermodynamic limit into question, and its microscopic origin remains unknown. Here, we show that this slow dynamics originates from many-body quantum beats that arise from the interaction-induced modulation of oscillations associated with single-particle hopping processes. We relate this mechanism to local integrals of motion (LIOMs) and show that it survives at arbitrarily large distances between LIOMs, is consistent with the stability of MBL in the thermodynamic limit, and is generic to many-body systems.
We present new numerical results for quasiperiodic potentials and, based on the quantum beats mechanism, develop an analytical model that explains number entropy growth and quasiparticle spreading, as well as clarifies the distinct MBL phenomenology observed in quasiperiodic and random potentials. Lastly, we propose concrete signatures for observing many-body quantum beats in existing experimental platforms.
\end{abstract}
\maketitle

One of the central questions of statistical physics is: Are there many-body
systems that do not thermalize and therefore cannot be described by a
statistical approach~\cite{Anderson1958,Anderson1979,Gornyi2005,Basko2006}?
Beyond fundamental significance, systems evading thermalization have
potential applications in emerging quantum technologies, where it is crucial
to avoid quantum decoherence. For a while, it was believed that many-body
localization (MBL)~\cite{Gornyi2005,Basko2006,Imbrie2016} provides a robust
mechanism to avoid thermalization in a closed
system~\cite{Abanin2019rev}. However, several recent results have called into
question the existence of a stable MBL phase (for a review,
see~\cite{Sierant2025rev}), although the MBL regime in which finite systems
avoid thermalization is indisputably present.

The aforementioned results include the numerical observation of slow particle dynamics probed by number entropy $S_{n}$~\cite{Kiefer2020}, defined as $S_n=-\sum_n p_{n, A}\ln p_{n, A}$, where $p_{n, A}$ is the probability of finding $n$ particles in subsystem $A$~\cite{Lukin2019}. In systems with a global conservation law, such as the conserved total particle number, entanglement entropy $S_e$ ---defined between the subsystem $A$ and the rest of the system--- can be split into number entropy $S_n$ and configurational entropy $S_c$ as $S_e = S_n+S_c$. As the transport of particles across boundaries of $A$ can be probed through $S_n$, it serves as a measure of localization, while its further appeal stems from the fact that it is an experimentally accessible quantity~\cite{Lukin2019}.

The absence of transport implies that $S_n$ saturates to a
sub-ergodic value~\cite{Bardarson2012, Singh2016, Luitz2020}. However, a
slow, and possibly unbounded, growth of $S_n$ observed even deep in the MBL regime and at timescales much longer than single-particle relaxation
time~\cite{Kiefer2020, Kiefer2021, Kiefer2021b, Kiefer2022,
Kiefer2022comment} suggests slow particle transport and a potential instability of the MBL phase in the thermodynamic limit. On the other hand, subsequent works~\cite{Luitz2020,Ghosh2022, Ghosh2022response} argued that the growth is due to local single-particle resonances, which do not destabilize the MBL phase. Later work observed a similar growth of $S_n$ in systems governed by a Hamiltonian explicitly constructed from local integrals of motion (LIOMs), and connected it to the slow spreading of the central quasiparticle~\cite{Chavez2024}. This suggests that the growth of $S_n$ does not necessarily imply the absence of the MBL phase, and that it is caused by a process that is, in some sense, local and generic to MBL systems. However, the microscopic mechanism underlying this process remains unknown, and the debate about the stability of the MBL in one dimension is still open.

Here, we find that the slow particle dynamics is caused by many-body quantum beats. Quantum beats are an interference effect arising when the system's state is in a superposition of energy eigenstates with small energy differences, leading to a beating pattern in intensity similar to acoustic beats~\cite{Forrester1955, Scully1997, Han2021}. The many-body analogue of this phenomenon emerges in MBL systems. When a near-resonant single-particle hopping occurs, single-site occupations exhibit Rabi-like oscillations, while weak interactions between two such hopping processes lead to interference and cause beats. From a phenomenological LIOM picture, which generally holds in MBL systems~\cite{Serbyn2013,Huse2014,Chandran2015}, a particle hopping between neighbouring sites can be described with two states of a LIOM, and the interaction between hopping processes is equivalent to the interaction between LIOMs.

We begin our study with the numerical investigation of MBL in quasiperiodic (QP) external potential which helps us to unveil and verify the quantum beats mechanism, as the deterministic nature of QP potential allows us to control the locations where resonances occur. Moreover, since there are no rare regions of weak potential, MBL seems to be more robust in QP than in random systems~\cite{Strkalj2021,Sierant2022, Strkalj2022, Falcao2024, Sierant2025rev}. There were indications that quasiperiodic systems exhibit $S_n$ growth in the MBL regime~\cite{Sierant2022} similarly to random systems, but detailed analysis and explanation was lacking. We focus on the half-chain number entropy $S_n$ and central quasiparticle width $\sigma_x^2$ in the XXZ Hamiltonian, and show that both quantities exhibit slow growth deep in the MBL regime, but with important qualitative differences relative to their random counterparts. 
We demonstrate that the proposed quantum beats mechanism provides a comprehensive account of the observed dynamics by developing an effective model that reproduces the exact numerical results. 
Furthermore, we verify the universality of our mechanism by showing its direct applicability to random systems. Finally, we show that many-body quantum beats can be directly observed and manipulated in the current state-of-the-art experiments, providing a direct experimental probe of interactions between LIOMs.

\begin{figure}[t!]
    \centering
    \includegraphics[scale=1]{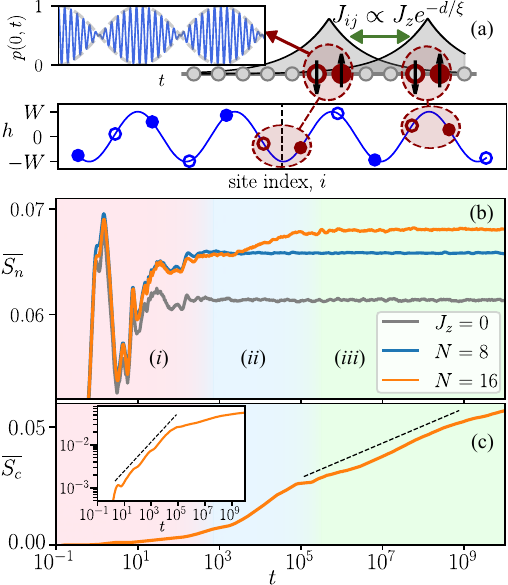}
    \caption{\textbf{Sketch of the proposed mechanism and the numerically calculated entropies.}
    \textbf{a}, Illustration of a chain with QP potential in the N\'{e}el
    state with filled/empty dots representing particles/holes. In the
    localized regime, dynamics is dominated by hopping between neighbouring
    sites with modest potential differences, circled in red. Quantum beats occur due to the modulation of the oscillation amplitude of single-site occupation probability (top left inset) caused by the interaction between near-resonant hopping processes. This interaction is equivalent to the interaction $J_{ij}$ between two LIOMs (top right). Dashed vertical line denotes the boundary between the subsystems relevant for the calculation of entropies.
    \textbf{b}, Moving time average of the sample-averaged number entropy
    $\overline{S_n}(t)$ for $W=6$ scaled with $(N-1)/N$, as explained in
    the Supplementary Information. Short and medium timescales (i) are shaded
    in pink, long timescales (ii) are shaded in blue and ultra-long timescales
    (iii) are shaded in green. The grey line is obtained for a non-interacting
    system ($J_z=0$) and $N=16$.
    \textbf{c}, Moving time average of the sample-averaged configurational
    entropy $\overline{S_c}(t)$ for $W=6$ and $N=16$. The inset shows the
    same curve on a log-log scale. Dashed lines denote the logarithmic and
    power-law growth at different time intervals in the main figure and the
    inset, respectively.}
    \label{fig1}
\end{figure}

\section{Model}

As a paradigmatic model for study of MBL, we consider a spin-1/2 chain described by the XXZ Hamiltonian
\begin{equation}
    H=J\sum_i \left( s_i^xs_{i+1}^x+s_i^ys_{i+1}^y \right)
     +J_z\sum_i s_i^z s_{i+1}^z
     + \sum_i h_i s_i^z \, ,
    \label{H_XXZ}
\end{equation}
with open boundary conditions and either QP external field given by the
Aubry-Andr\'{e} potential~\cite{Aubry1980}
$h_i=W\cos(2\pi\beta i+\phi)$, $\beta\in \mathbb{R} \setminus \mathbb{Q}$ (see Fig.~\ref{fig1}a), or random disorder $h_i\in [-W, W]$ drawn from uniform distribution. The units are fixed by taking $\hbar=1$ and $J=-1$. We also take $J_z=-1$ (isotropic Heisenberg chain) and set the spatial frequency to the inverse of the golden mean, $\beta=2/(1+\sqrt{5})$, unless indicated otherwise. The Jordan-Wigner transformation~\cite{Jordan1928} maps the Hamiltonian~\eqref{H_XXZ} to an equivalent fermionic model with conserved particle number; hence, we interchangeably use the terms ``particle'' and ``spin'' excitation.

\section{Growth of the number entropy}

Our numerical result for $S_n$ in QP chains is shown in Fig.~\ref{fig1}b. We observe that the dynamics of $S_n(t)$ is not frozen at long timescales of up to $t=10^5$ tunnelling times (i.e.\ $|J|^{-1}$) when the system is in the MBL regime. By comparison, the non-interacting dynamics is already fully saturated at $t<10^3$. In contrast to the random potential, $\overline{S_n}(t)$ shows structured behaviour at long timescales instead of featureless growth that can be fitted as $\log\log t$ or a power-law discussed in
Refs.~\cite{Kiefer2020, Kiefer2021, Ghosh2022}.

In Fig.~\ref{fig1}b, we identify three regions in the behaviour of
$\overline{S_n}(t)$, that correspond to (\textit{i}) short and medium,
(\textit{ii}) long and (\textit{iii}) ultra-long timescales. In region
(\textit{i}), $\overline{S_n}(t)$ is dominated by the single-particle
dynamics without qualitative differences between interacting an non-interacting chains, and with all curves being independent of $N$. On the other hand, for long times in (\textit{ii}), we observe a striking difference: for $N\leq 8$ the dynamics of interacting chains is saturated as in the non-interacting case, while for $10\leq N\leq 16$ the averaged number entropy $\overline{S_n}$ exhibits further growth, saturating only much later, around $t\sim 10^5$ independently of $N$ (see also Fig.~\ref{fig3}b). This indicates the many-body nature of the processes responsible for the growth. At ultra-long timescales in (\textit{iii}), we observe the size-dependent plateau, consistent with the lack of transport in the MBL regime.

Concurrently, we show the averaged configurational entropy $\overline{S_c}$
in Fig.~\ref{fig1}c, and observe an algebraic growth in (\textit{i}) and
(\textit{ii}), indicating weakly ergodic behaviour. In region (\textit{iii}),
we observe the logarithmic growth, as expected in the MBL regime.
Interestingly, we observe no saturation of $\overline{S_c}$ all the way up
to $t\sim10^9$ ---orders of magnitude after $\overline{S_n}$ is saturated---
implying that the plateau in $\overline{S_n}$ is a feature of many-body
dynamics, and does not come from finite-size effects. Lastly, we have checked
that the qualitative features of our results do not depend on $\beta$, $J_z$, $W$ (as long as the system is in the MBL regime), nor the choice of particular
initial state, c.f.\ Figs.~\ref{fig2} and~\ref{fig3}.

\section{Many-body quantum beats}

\begin{figure}[t!]
    \centering
    \includegraphics[scale=1]{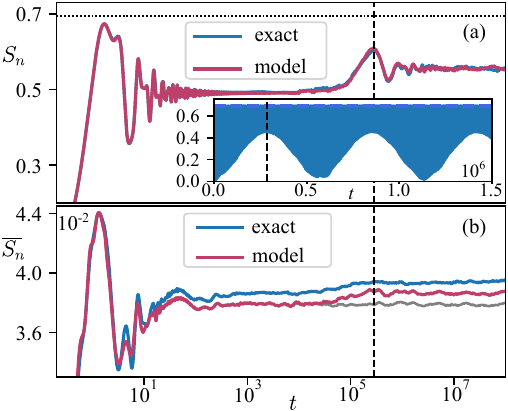}
    \caption{\textbf{Comparison of exact dynamics with the results obtained
    from our effective model.}
    \textbf{a}, A moving time average of $S_n(t)$ for a single configuration
    of QP potential with $\phi = \phi_{\rm PR}=3.863$ (same as in
    Fig.~\ref{fig1}a and corresponding to the point-reflection symmetry
    position between $i=7$ and $i=8$ ---left-most site has index $i=0$) and
    comparison with effective model from Eq.~\eqref{Ham4}. We used the initial
    N\'{e}el state, $N=12$ and $W=10$. The inset shows the same $S_n(t)$
    obtained from ED, without time averaging and with a linear time scale. A
    dashed vertical line marks the onset of beats at $t=\pi/\epsilon$. The
    dotted horizontal line corresponds to $\ln 2$.
    \textbf{b}, A comparison of moving time average of $\overline{S_n}(t)$
    for $W=10$ obtained from ED with $N=16$ sites and from our effective
    model. The grey line shows the results for single-particle dynamics
    described by Eq.~\eqref{S_n_sum}.}
    \label{fig2}
\end{figure}

As we discuss in the following, the observed long-time dynamics of $S_n$,
i.e.\ (\textit{ii}) and (\textit{iii}) in Fig.~\ref{fig1}b, is fully captured
by the mechanism shown in Fig.~\ref{fig1}a: the amplitude of single-particle
Rabi-like oscillations between two sites is modulated due to the interference with other nearly resonant single-particle hoppings (predominantly nearest-neighbour). To see how this modulation occurs, let us consider four sites $a$, $b$, $c$ and $d$, e.g.\ red sites in Fig.~\ref{fig1}a, and the subspace of Hilbert space spanned with
$\{\ket{1}\equiv\ket{1_a0_b1_c0_d},\; \ket{2}\equiv\ket{1_a0_b0_c1_d},\;
\ket{3}\equiv\ket{0_a1_b1_c0_d},\; \ket{4}\equiv\ket{0_a1_b0_c1_d}\}$,
where $1_a$/$0_a$ denotes presence/absence of the particle at site $a$.
Hamiltonian~\eqref{H_XXZ} restricted to this subspace can be written as
\begin{equation}
    \label{Ham4}
    H_{\rm AM}= \begin{psmallmatrix}
        E_1^0 & J/2 & J/2 & 0\\
        J/2 & E_2^0 & 0 & J/2\\
        J/2 & 0 & E_3^0 & J/2\\
        0 & J/2 & J/2 & E_4^0
    \end{psmallmatrix} \, .
\end{equation}

Here, $E_i^0=\bra{i}H\ket{i}$ are unperturbed energies in state $\ket{i}$.
Eigenvalues of the Hamiltonian above come in two pairs so that
$E_1+E_4=E_2+E_3=\overline{E}=(E_1+E_2+E_3+E_4)/4$. Without loss of
generality, we choose $\overline{E}=0$, after which it follows that $E_1=-E_4$,
$E_2=-E_3$. If the differences between diagonal elements of $H_{\rm AM}$ are
small compared to $J$, eigenvalues and eigenvectors are approximately given by:
$E_1= J-A$, $\ket{v_1}\approx(1/2,1/2,1/2,1/2)$;
$E_2=-A$, $\ket{v_2}\approx(1/2,1/2,-1/2,-1/2)$;
$E_3=A$, $\ket{v_3}\approx(1/2,-1/2,1/2,-1/2)$;
$E_4=-J+A$, $\ket{v_4}\approx(1/2,-1/2,-1/2,1/2)$,
where the exact form of $A$ is irrelevant for the following discussion.
If relations $E_1=-E_4$ and $E_2=-E_3$ held exactly, there would be no amplitude modulation (AM) and
the number entropy would not grow, as is the case in non-interacting systems.
However, the presence of interactions enables the rest of the system to act as
a mediator between two single-particle hopping processes. Consequently, small
effective corrections to the eigenvalues of $H_{\rm AM}$ arise from the change
of the effective field at locations $a$ and $b$ when a particle jumps from $c$
to $d$, and vice versa.

The discussion above naturally connects to the LIOM
description~\cite{Serbyn2013,Huse2014,Chandran2015}. Near-resonant
single-particle hoppings give rise to two LIOMs $\tau_{ab}^z$ and
$\tau_{cd}^z$, centred on sites $a, b$ and $c, d$, respectively, and
exponentially decaying away from them. They have approximate eigenstates
$\ket{\pm}_{ij}\approx (\ket{01}_{ij}\pm \ket{10}_{ij})/\sqrt{2}$ and can be
described by the Hamiltonian
$H_{\rm LIOM}=\tilde{h}_{ab}\tau_{ab}^z+\tilde{h}_{cd}\tau_{cd}^z
+J_{ab,cd}\tau_{ab}^z\tau_{cd}^z$ to the lowest order. The last term captures
the interaction arising from the effective field change mentioned above, where
$J_{ab,cd}\propto J_ze^{-d/\xi}$ to lowest order~\cite{Huse2014}. Going back
to the $H_{\rm AM}$ eigenbasis, we identify the LIOM eigenstates as
$\ket{++}=\ket{v_1}$, $\ket{-+}=\ket{v_2}$, $\ket{+-}=\ket{v_3}$ and
$\ket{--}=\ket{v_4}$, with shifted eigenvalues: $E_{1,4}\to E_{1,4}+J_{ab,cd}$
and $E_{2,3}\to E_{2,3}-J_{ab,cd}$, that lead to AM and beats.

Lastly, note that unlike the resonance-based mechanisms~\cite{Ghosh2022}, the
quantum beats mechanism requires no narrow resonances, relying instead on interactions between single-particle processes.

To be specific, let us take the whole system to be initially in state
$\ket{4}$. This choice does not qualitatively affect any of our conclusions.
When the boundary between the subsystems stands between sites $a$ and $b$, as
in Fig.~\ref{fig1}a, we have $S_n(t)=-p(0,t)\ln p(0,t)-p(1,t)\ln p(1, t)$.
Probabilities $p(1,t)=1-p(0,t)$, $p(0,t)=p(\ket{3}, t)+p(\ket{4},t)$
correspond to having one or zero particles in the subsystem on the left,
respectively. Note that, at this moment, we treat other particles in the chain
as being frozen and therefore having no effect on the dynamics of two particles
on sites $a, b, c, d$.
From $\ket{\psi(t)}=\sum_{j=1}^4 \ket{v_j}\langle v_j\ket{4}\exp(-iE_jt)$
and analytical forms of $E_j$ and $\ket{v_j}$ written above, we obtain
\begin{equation}
    p(0,t)=\frac{1}{2}\Bigg[1+\cos\!\left(\frac{\epsilon}{2}t\right)
    \cos\!\Big(\!\left(1+\frac{\epsilon}{2}\right)t\Big) \Bigg] \, ,
    \label{p0beats}
\end{equation}
with $\epsilon=4J_{ab, cd}$. The term $\cos(\frac{\epsilon}{2}t)$ modulates
the oscillation amplitude on long timescales and causes quantum beats.

From Fig.~\ref{fig2}a, we observe excellent agreement of $S_n$ obtained from
our effective model with the exact numerical calculation for $N=12$ and a
particular phase $\phi$ of the QP potential. The parameter $\epsilon$ is taken from numerical eigenvalues as $\epsilon=E_1+E_4-E_2-E_3$. The timescale at which beats occur is set by $1/\epsilon$, which diverges as the energy difference vanishes. Beats lead to an increase of the time average
$\langle S_n(t)\rangle=1/t\int_0^t S_n(t') \mathrm{d} t'$, as we show by using
the definition of $S_n$ and employing Eq.~\eqref{p0beats} to obtain
$\langle S_n(t)\rangle\approx 0.39$ for $t\ll \epsilon^{-1}$ and
$\langle S_n(t)\rangle\approx 0.55$ for $t\gg \epsilon^{-1}$. Intuitively, suppression of oscillation amplitude causes the particle oscillating across the boundary to spend more time delocalized between sites, resulting in an increased $\langle S_n(t)\rangle$.

The reason why this mechanism captures the long-time growth of $S_n$ so well is because $\epsilon\propto J_{ab,cd}$ is exponentially small in the distance $d$ between pairs of sites with strong hybridization. The rapid growth of timescale with the distance $d$ between the pairs is visible from Fig.~\ref{fig3}a, where we show $\overline{S_n}(t)$ for different $d$, which are tuned by changing the QP frequency $\beta$. However, in contrast to noninteracting, single-particle resonances, there is no suppression of effective tunnelling matrix elements as the elements of the effective Hamiltonian~\eqref{Ham4} do not depend on distance $d$. Therefore, in a sufficiently large system, every single-particle hopping process will eventually find its pair so that differences between diagonal elements of the Hamiltonian~\eqref{Ham4} are small enough, leading to significant AM at long enough timescales.

\section{Validity of the effective model}

\begin{figure}
    \centering
    \includegraphics[scale=1]{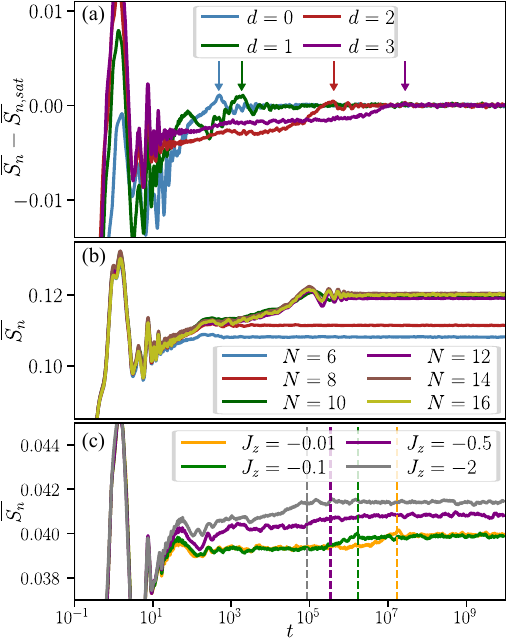}
    \caption{\textbf{Dependence of saturation timescales on distance $d$, chain length $N$ and interaction strength $J_z$.}
    \textbf{a}, Moving time average of $\overline{S_n}(t)$ for initial N\'{e}el state, $N=12$, $W=10$ and different frequencies $\beta$ corresponding to $d=0$ for $\beta=13/50$, $d=1$ for $\beta=8/25$, $d=2$ for
    $\beta=2/(1+\sqrt{5})$ and $d=3$ for $\beta=1/\sqrt{2}$. Arrows mark
    saturation times for different $d$. Note that the choice of a rational
    $\beta=p/q$ is numerically indistinguishable from the irrational case for
    $q>N$. $(\cdot)_{\rm sat}$ denotes the saturation value calculated
    numerically as the average of a given quantity for $t>10^8$. 
    \textbf{b}, Moving time average of $\overline{S_n}(t)$ for initial N\'{e}el state, $W=6$, $\beta=2/(1+\sqrt{5})$ and different chain lengths $N$.
    \textbf{c}, Moving average of $\overline{S_n}(t)$ for $W=10$, $N=12$, $\beta=2/(1+\sqrt{5})$ and different $J_z$. Dashed vertical lines show the analytical prediction of linear scaling for the saturation time with $J_z$.}
    \label{fig3}
\end{figure}

In Fig.~\ref{fig2}b, we show excellent agreement between our combined
effective model, which takes into account both single-particle hoppings and AM,
and the numerical results obtained by exact diagonalization (ED). The procedure used to combine the
models and obtain $S_n$ averages is described in the Supplementary
Information. Briefly, the averaging is done by integrating QP field
configurations over $\phi\in[0, 2\pi]$. The only external parameter is the correction $\epsilon$ that determines the beat timescale. As discussed previously, $\epsilon$ is set by distance $d$, which is in turn fixed by QP field frequency $\beta$. In practice, we choose the value of $\epsilon$ so that it gives the same beat timescale as the one obtained for a single phase $\phi_{\rm PR}$ of point reflection symmetry, see Figs.~\ref{fig1}a and~\ref{fig2}a.

Results in Fig.~\ref{fig3}b and~\ref{fig3}c give further support to our model, showing that the $S_n$ growth in QP systems is dominated by beats at a particular $d$ independently of chain length $N$, and that the saturation timescales are proportional to $J_z^{-1}$, in line with the prediction from the LIOM interaction strength. We can also explain the difference in saturation times of $S_n$ and $S_c$ in QP chains, see Figs.~\ref{fig1}b and~\ref{fig1}c. The saturation of $S_n$ is governed by appearance of beats at a timescale $1/\epsilon \propto e^d$, while relevant timescales for the growth of $S_c$ are proportional to $e^{N/2}$, when the central and edge sites entangle and $S_c$ starts deviating from logarithmic growth, and $e^N$, when the entanglement spreads from edge to edge, therefore completely saturating $S_c$. In QP chains $d$ is fixed by $\beta$ and can be smaller than $N/2$, $S_n$ can saturate while $S_c$ still grows logarithmically. Moreover, for chains that are short enough, beats do not appear at all, leading to much earlier saturation of $S_n$, see the $N=8$ curve in Fig.~\ref{fig1}b. On the other hand, in the random system, distance $d$ is also random, so interference between hoppings on central and edge sites is possible, which explains the previous observation that saturation of $S_n$ and $S_c$ starts occurring at the same timescale~\cite{Kiefer2021}.

\begin{figure}
    \centering
    \includegraphics[scale=1.]{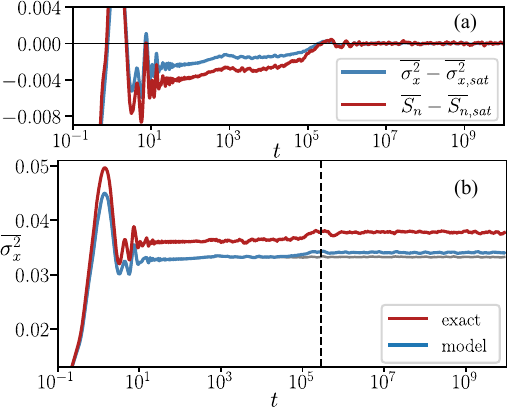}
    \caption{\textbf{Quasiparticle spreading.}
    \textbf{a}, Moving time average of $\overline{\sigma_x^2}(t)$ compared with $\overline{S_n}(t)$ for $W=6$, $N=12$, $\beta=2/(1+\sqrt{5})$ and initial N\'{e}el state.
    \textbf{b}, Comparison of moving time average of $\overline{\sigma_x^2}(t)$ obtained from ED and from our effective model, for $W=10$, $N=12$ and the initial N\'{e}el state. Grey line denotes the results for a non-interacting ($J_z=0$) system.
    }
    \label{fig4}
\end{figure}

To extend the scope of our work beyond $S_n$, we study the central
quasiparticle width $\sigma_x^2$, see Methods for the definition. We show in Fig.~\ref{fig4}a that the time evolutions of $\sigma_x^2$ and $S_n$ are closely related, sharing all qualitative features. Our effective model also captures the behaviour of quasiparticle width as shown in Fig.~\ref{fig4}b. This further supports the view that the growth of $S_n$ is a consequence of a local process, specifically quantum beats. It also shows that quantum beats are a generic phenomenon in many-body systems that can explain the behaviour of various dynamical quantities. By showing the connection between $S_n$ and $\sigma_x^2$ behaviour in a system
governed by the QP XXZ model Hamiltonian, we expand on the
conclusions from Ref.~\cite{Chavez2024}, which related these quantities when
the Hamiltonian is explicitly constructed from LIOMs.

\section{Random disorder and experimental realization}

\begin{figure}[ht!]
    \centering
    \includegraphics[scale=1.]{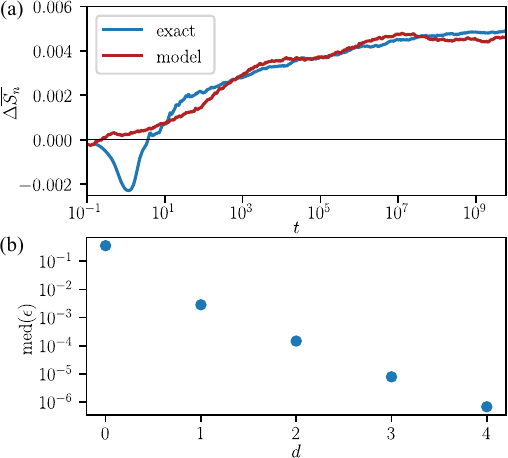}
    \caption{\textbf{Results with random disorder.}
    \textbf{a}, Growth of $S_n$ induced by interactions for a system with random disorder, $W=15$ and $N=16$. The model results are plotted with the rescaled time $t'=t^d_{\rm max}/T$ as explained in the main text. For $N=16$ we have $d_{\rm max}=5$, while the fit parameters are $T=1.38\times 10^{22}$ and $A=3.98$, in line with the expectation $A\sim d_{\rm max}$.
    \textbf{b}, Median of $\epsilon=E_1+E_4-E_2-E_3$ calculated from configurations with near-resonant pairs implanted with a distance $d$ between them. We use random disorder with independent variables $h_i\in[-W, W]$ drawn from a uniform distribution, $W=20$ and chain length $N=10$. Near-resonant pairs are implanted so that $h_1$, $h_2$, $h_{d+3}$ and $h_{d+4}$ are chosen from the uniform distribution in the interval $[-1,1]$.
    }
    \label{fig5}
\end{figure}

As discussed above, quantum beats are a generic mechanism that applies to QP and random systems in a conceptually identical way. The only difference is that in random systems $d$ is not fixed, so the distribution of timescales $\epsilon^{-1}$ is much wider. 
In what follows, we utilize our effective model based on Eq.~\eqref{Ham4} in a random external field, and compare its predictions with exact numerics (Fig.~\ref{fig5}a) to demonstrate the validity of our approach in random systems. 
We isolate the growth $\Delta S_n$ caused by beats by subtracting the $S_n$ average for $t\ll \epsilon^{-1}$ (see Supplementary Information~\ref{sec:effmodel}), and compare it to the $S_n$ growth caused by interaction effects, captured by the difference $\Delta \overline{S_n}_{, \,\rm exact}(t)=\overline{S_n}(t, J_z=-1)-\overline{S_n}(t, J_z=0)$ obtained from exact numerics. 

To capture the wide distribution of timescales, we assume that random $d$ gives rise to two effects: exponential widening of the growth time window and enhanced growth magnitude compared to the case with a single allowed $d$.
Technically, we fit the curve $A\cdot\Delta \overline{S_n}_{,\,\rm model}(t')$ to the exact result $\Delta \overline{S_n}_{, \,\rm exact}(t)$. The rescaled time is defined as $t'=t^{d_{\rm max}}/T$, where $d_{\rm max}$ is the maximum $d$ allowed by the system size, so the only fit parameters are $A$ and $T$. Physically, we expect $A\sim d_{\rm max}$ since this gives the number of possibilities where beats can occur, while $T$ determines the timescale at which beats start occurring. The agreement between the results of our simple model and exact numerics in Fig.~\ref{fig5}a shows that quantum beats can explain $S_n$ growth in random MBL systems.

\begin{figure}[t!]
    \centering
    \includegraphics[scale=1.]{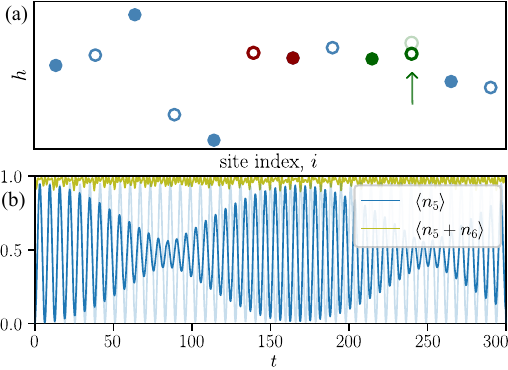}
    \caption{\textbf{Field configurations and corresponding occupation probabilities for the proposed experimental observation of the quantum beats.} 
    \textbf{a}, The field configuration used to obtain beats at experimentally accessible timescales. The red and green sites are near-resonant, we use $h_5=5.1, h_6=2.2, h_7=8, h_8=1.7, h_9=4.6$, and $h_i\in [-50, 50]$  for other sites.
     Transparent green dot marks the displaced field strength $h_9$, for which the green sites are no longer near-resonant.
    \textbf{b}, AM on a timescale shorter than $10^2$ tunnelling times for the field configuration shown in \textbf{a}, with $|J_z| = 3$ and the N\'{e}el initial state. The occupation number at site $j=5$ (the empty red site in \textbf{a}) shows Rabi-like oscillations with beats (dark blue curve), while the sum $\langle n_5+n_6\rangle\approx 1$ confirms that the particle remains localized on the red sites.
    The transparent blue curve shows the result for the configuration in \textbf{a} with the displaced field $h_9$, for which the quantum beats disappear.
    }
    \label{fig6}
\end{figure}

Let us discuss a viable proposal for the experimental observation of quantum beats in MBL systems. To this end, we explore direct manipulation of quantum beats by implanting pairs of sites with weak (or no) disorder, while keeping the rest of the system strongly localized.
By choosing to have weak disorder on pairs of sites separated by $d$, in Fig.~\ref{fig5}b we directly show the exponential decay of $\epsilon$ with $d$, in accordance with the exponential decay of interactions between two LIOMs. 

Such implanted configurations can be readily implemented in cold atom platforms where MBL signatures have already been firmly established~\cite{Lukin2019,Leonard2023}. The key challenge is having beats on experimentally accessible timescale, i.e. a few hundred tunnelling times. To achieve this, one can use a configuration with $d=1$ and moderately increased interaction strength $J_z=-3$, see Fig.~\ref{fig6}a. In this case, the beats occur at $t\sim 100$ (experimentally accessible) tunnelling times, see Fig.~\ref{fig6}b, corresponding to a few tens of Rabi-like oscillation periods. By measuring the occupation of the target site during this time period and observing the collapse and revival of oscillations, quantum beats could be observed directly. Furthermore, by slightly tuning the field strength $h_i$ at one site in one pair, the pair becomes non-resonant which immediately destroys the beats in the other, resonant pair. This confirms that the interference between near-resonant hoppings is directly responsible for the quantum beats.

\section{Conclusion and outlook}

The significance of our findings is twofold: (i) our observations and the
effective model expand the understanding of the MBL regime, explaining the slow
many-body dynamics that does not involve long-range transport and is consistent with the stability of MBL phase; (ii) we found that quantum beats, a well-known phenomenon in atomic, molecular and solid-state physics, are important for the dynamics in MBL systems and seem relevant in a wider context of quantum many-body systems. Quantum beats may also provide a microscopic explanation for the long-range correlations near the MBL transition reported in Refs.~\cite{Laflorencie2025,Padhan2026}. Experimental relevance of our work extends beyond the proposed observation of quantum beats in MBL systems, since measurement of beat timescales is also a direct probe of the interactions between LIOMs, and could be used to test phenomenological descriptions of MBL. Finally, our work highlights the rich phenomenology that arises from interactions in localized systems, where non-ergodicity does not necessarily imply frozen dynamics.

\section{Methods}

\subsection{Numerical details}
All numerical results are obtained from full exact diagonalization (ED) of the Hamiltonian~\eqref{H_XXZ}. Observables are averaged over potential
realizations obtained by randomly sampling the phase $\phi \in [0, 2\pi]$ or $h_i\in[-W, W]$ for random disorder, and the initial configurational basis states from the half-filling sector, except when indicated that fixed initial N\'{e}el state is used. Additionally, unless indicated otherwise, moving time averages are computed so that we can discern trends in dynamics more easily. To focus on the genuine features of MBL dynamics and avoid crossover effects, we set $W\geq 6$ to ensure that the system is in a deeply localized regime~\cite{Iyer2013, Doggen2019, Singh2021, Sierant2022}.

\subsection{Single-particle dynamics}
Our description of the single-particle dynamics is based on the generalized version of the approach from Ref.~\cite{Ghosh2022}. Namely, we consider single-particle hoppings with a range of 1 site (nearest-neighbour hoppings), 2 sites (next-nearest-neighbour hoppings) and further. In general, a single-particle hopping of range $k$ between sites $j$ and $j+k$ can be described with an effective two-dimensional Hamiltonian
\def\B{
\begin{psmallmatrix}
    E_1^0 & j_{\rm eff}\\
    j_{\rm eff} & E_2^0
\end{psmallmatrix}}
\begin{equation}
    H_{\rm res}=\B \, ,
    \label{Ham2}
\end{equation}
in the subspace spanned with
$\ket{\phi_1}=\ket{\dots 0_j\dots 1_{j+k}\dots}$ and
$\ket{\phi_2}=\ket{\dots 1_j\dots 0_{j+k}\dots}$. Again, $E_i^0$ denotes
unperturbed energies, and $j_{\rm eff}(k)\sim W^{-k+1}$ denotes the effective
tunnelling element that decays exponentially with distance $k$.

Let us take, without loss of generality, $\ket{\phi_1}$ as the initial state.
Time evolution with $H_{\rm res}$ in Eq.~\eqref{Ham2} allows us to calculate
the number entropy $S_n=-P_1(t)\ln P_1(t)-P_2(t)\ln P_2(t)$, where
$P_1 \equiv P(\ket{\phi_1})$, $P_2 \equiv P(\ket{\phi_2})=1-P_1$, and we
suppose that the particle hops across the boundary. We obtain
\begin{equation}
    \overline{S}_{n, {\rm res}}(t)=f\frac{1}{2\pi}\int_{0}^{2\pi}
    \mathrm{d}\phi \; S_n(\phi, t) \, ,
    \label{S_n_int}
\end{equation}
where $S_n(\phi, t)$ is calculated as explained above with a phase $\phi$ used
to determine $h_j$ and $h_{j+k}$ which fixes the matrix elements in
Eq.~\eqref{Ham2}. The origin of the combinatorial factor $f$ is explained in
the Supplementary Information. It is enough to apply Eq.~\eqref{S_n_int} to
range $k=1$ hoppings, after which generalization to all single-particle
hoppings follows directly:
\begin{equation}
    \overline{S}_{n, {\rm sp}}(t)=\sum_{k=1}^{\infty}k\left|\frac{j_{\rm eff}(k)}
    {j_{\rm eff}(1)}\right|\overline{S}_{n, {\rm res}}\!\left(\left|\frac{j_{\rm eff}(k)}
    {j_{\rm eff}(1)}\right|t\right) \, .
    \label{S_n_sum}
\end{equation}
Note that the first factor $k$ in the sum describes the possibility of range
$k$ cross-boundary hopping to occur at $k$ positions.

\subsection{Quasiparticle dynamics}

We define the quasiparticle width in the same way as in
Ref.~\cite{Chavez2024}. Let us consider a subsystem $A$ comprised of
the first $k$ sites counting from the left, i.e.\ sites $i=0$ to $i=k-1$.
The probability of finding $n$ particles in the subsystem $A$ is
\begin{equation}
    p_{n, A}=p_n(k)=\sum_{\langle a|N_A\ket{a}=n}
    \lvert\langle a\ket{\psi(t)} \rvert^2,
\end{equation}
where $N_A$ is the particle number operator for the subsystem $A$.
Probabilities $p_{n}(k)$ now enable us to calculate the probability of the
particle number growing from $n-1$ to $n$ at a site $i$~\cite{Chavez2024}:
\begin{equation}
    q_i(n)=\sum_{m=n}^{N}\Big( p_i(m)- p_{i-1}(m)\Big),
\end{equation}
where we set $n\geq 1$ and $p_{-1}(m)=0$. Since $\sum_{i=0}^{N-1}q_i(n)=1$
and $\sum_{n=1}^N q_i(n)=\langle n_i\rangle$, $q_i(n)$ can be viewed as a
probability distribution for the position of the $n$-th quasiparticle. The
mean and the variance of this distribution are given by
\begin{align}
    \langle x_n\rangle &=\sum_{i=0}^{N-1} i \, q_i(n) \, ,\\ \nonumber
    \sigma_{x_n}^2 &=\sum_{i=0}^{N-1} \big( i-\langle x_n\rangle \big)^2 q_i(n) \, .
\end{align}
As in Ref.~\cite{Chavez2024}, our results are obtained by focusing on the
central quasiparticle, i.e.\ $\sigma_{x_{N/4}}^2\equiv \sigma_x^2$.
Our effective model can straightforwardly be used to predict the behaviour of the sample-averaged quasiparticle width, $\overline{\sigma_x^2}$. This is achieved by restricting the quasiparticle support on two sites $a$ and $b$. In the quantum beat model from Eq.~\eqref{Ham4}, these correspond to one pair of the
nearly-resonant sites (see Fig.~\ref{fig1}a and the explanation in the main
text), and the quasiparticle width $\sigma_x^2$ is calculated from the
distance between $a$ and $b$ and probabilities $p(0, t)$ and $p(1, t)$ of
finding zero or one particle in a given subsystem, respectively. In the
single-particle resonance model of Eq.~\eqref{Ham2}, the same procedure is
used, with $a$ and $b$ denoting the near-resonant sites. Averages are obtained
in a way conceptually identical to the one described in Supplementary
Section~\ref{sec:effmodel} for number entropy.

%
\section{Acknowledgements}
%
We thank Sebastian Schmid for fruitful discussions during the early phase of this work, and Nicolas Laflorencie for useful comments.
The work of A.\v{S}. is supported by the European Union’s Horizon Europe research and innovation programme under the Marie Sk\l{}odowska-Curie Actions Grant agreement No. 101104378.
H.B. acknowledges support from the project “Implementation of cutting-edge research and its application as part of the Scientific Center of Excellence for Quantum and Complex Systems, and Representations of Lie Algebras”, Grant No. PK.1.1.10.0004, co-financed by the European Union through the European Regional Development Fund—Competitiveness and Cohesion Programme 2021-2027.
All data that support the plots within this paper are available upon request.

\bibliography{ref.bib}

\clearpage

\onecolumngrid
\setcounter{secnumdepth}{3}
\renewcommand{\thesection}{S.\arabic{section}}
\setcounter{section}{0}
\renewcommand{\theequation}{S.\arabic{equation}}
\setcounter{equation}{0}
\renewcommand{\thefigure}{S.\arabic{figure}}
\setcounter{figure}{0}
\renewcommand{\theHfigure}{S.\Alph{section}\arabic{figure}}
\renewcommand{\theHequation}{S.\Alph{section}\arabic{equation}}

\begin{center}
    \Large\textbf{Supplementary Information for\\ ``Quantum Beats in Many-Body Localized Systems''}\\
\end{center}
\begin{center}
Bernard Faulend$^*$, Hrvoje Buljan, Antonio \v{S}trkalj$^{\dagger}$

\mbox{Department of Physics, Faculty of Science, University of Zagreb, Bijeni\v{c}ka c. 32, 10000 Zagreb, Croatia}
\end{center}

\section{Higher-order quantum beats}
As discussed in the main text, the dynamics of QP MBL systems at long
timescales can be explained through quantum beats. The key point is that
the QP structure of the potential fixes the distance $d$ between pairs of
sites that are strongly hybridized. Quantum beats, caused by the interaction
between longer-range single-particle hoppings are also present (e.g. beats due to interaction between a range $k=2$ single-particle hopping and range
$k=3$ single-particle hopping). This makes the description more complex since we now have a set of distances $d$ for each combination of single-particle hopping ranges $k$. However, the sum of these contributions to $S_n$ is by a factor of order $W$ smaller than the main contribution, where both hoppings are nearest-neighbour. Therefore, it is justified to neglect these effects in our effective model, which, as we observe, does not adversely affect the agreement with the numerical results.

Furthermore, in large chains, additional pairs of sites with similar field
strengths exist, and therefore several pairs of sites hybridize. This leads to
multiple beating timescales, so we expect a series of step-like increases in the time average of $S_n$, with the growth --- at least the growth related to beats--- bounded from above by $S_n\leq\ln 2$, which corresponds to a single
particle oscillating across the boundary.

\section{Further details of numerical calculations}

In our numerical calculations, we first choose the initial state $\ket{\psi_0}$
and QP potential phase $\phi$, and then perform full exact diagonalization of
the Hamiltonian~\eqref{H_XXZ}. We construct the Hamiltonian using the QuSpin
Python package~\cite{quspin}. After obtaining the eigenvalues and eigenstates
of~\eqref{H_XXZ}, we time-evolve the initial state to get $\ket{\psi(t)}$ and
then calculate all quantities of interest such as $S_n$. Some of the key
features of our results appear only at very long evolution times which are out
of reach for all approximative methods. In practice, full ED is still limited
by the numerical precision of floating-point numbers, but using standard 64-bit
float allows us to study evolution times up to $t\sim 10^{12}$, which shows to
be enough to capture and explain all relevant effects. Averages in our plots
were taken over at least $2\times 10^4$ configurations, while moving time
averages were obtained by averaging over 50--200 logarithmically spaced time
points.

\section{Combinatorial corrections in a half-filling sector}

Averaging number entropy $S_n$ over computational basis states of the
half-filling sector results in non-monotonous dependence of saturation values
$\overline{S}_{n, \rm sat}$ on chain length $N$~\cite{Kiefer2021, Ghosh2022}.
This is due to the fact that the share of computational basis states with
allowed range $k$ single-particle hoppings decreases with $N$. While $S_n$ at
long timescales is larger in longer chains, short-time dynamics is dominated
by effectively single-particle resonances, leading to non-monotonous behaviour.
We define the combinatorial factor $f$ as the ratio of the number of
computational basis states with allowed range $k$ resonance to the total number
of computational basis states in the half-filling sector. Resonances are not
allowed if sites $j$ and $j+k$ are initially both filled or empty. In the
thermodynamic limit, $f=1/2$, while for finite systems
$f=N/(2N-2)$~\cite{Kiefer2022}. To cancel out this combinatorial effect, the
results can be scaled with $(N-1)/N$ so that short-time dynamics of
$\overline{S}_n$ for different $N$ overlap perfectly, and non-monotonous
behaviour disappears, see Fig.~\ref{fig1}b in the main text.

\section{Effective model for the number entropy\label{sec:effmodel}}

\begin{figure*}[h]
    \centering
    \includegraphics[scale=1.]{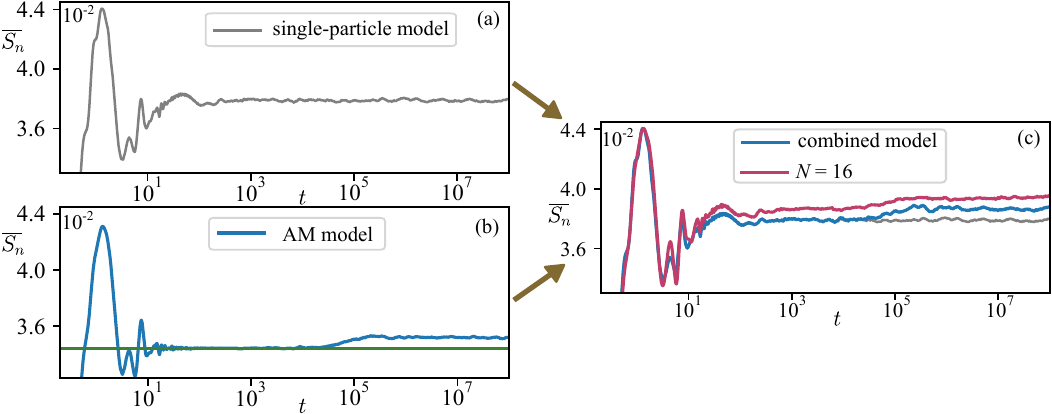}
    \caption{Moving time average of $\overline{S_n}(t)$ obtained from the
    single-particle model described by Eq.~\eqref{S_n_sum} (\textbf{a}), and
    from the AM model as described in the text (\textbf{b}). A horizontal
    green line denotes the time average of $\overline{S_n}(t)$ for
    $t\in[10^2, 5\times 10^3]$. Growth of $\overline{S_n}(t)$ for
    $t>5\times 10^3$ is ascribed to AM effects and added to single-particle
    results in the combined model. \textbf{c}, Comparison of the combined
    model and results of full ED, same as Fig.~\ref{fig2}b in the main text.
    Results in this figure were obtained for $W=10$ and sample averaging over
    both phase $\phi$ and initial state.}
    \label{smfig1}
\end{figure*}

In this section, we explain how we combined the analytical single-particle
model~\eqref{Ham2} and quantum beats model~\eqref{Ham4}, valid on short and long timescales respectively, into the effective model that qualitatively describes the dynamics over all accessible timescales. Number entropy average is obtained by integrating QP field configurations over a randomly chosen
$\phi \in [0,2\pi]$. We further take into account the combinatorial factor $f$
if averaging over initial states is done. Averaging over $\phi$ simulates the
results obtained for a random position in a very long chain. Technically, $\phi$ enters the effective Hamiltonians~\eqref{Ham4} and~\eqref{Ham2} through
diagonal elements $E_i^0$ that are determined by the potential energy in
external field $h_j$ and interaction energy of the configurational basis state
$\ket{i}$.

Single-particle contribution to $S_n$ is calculated from Eq.~\eqref{S_n_sum}
in the main text. Effective tunneling elements $j_{\rm eff}(k)$ that enter
Eq.~\eqref{S_n_sum} can be approximated with $j_{\rm eff, 0}(k)$ calculated
for the phase $\phi$ of a mirror-symmetry point. Resonances with even range $k$ occur around mirror-symmetry points $\phi=0, \pi$ where the field strengths satisfy $h_n=h_{-n}$ ($n\in\mathbb{Z}$), while resonances with odd range occur around $\phi=-\pi\beta, -\pi\beta+\pi$ for which $h_n=h_{-n+1}$.

To calculate $j_{\rm eff,0}(k)$, we note that in a single-particle system
energy eigenstates will be given approximately with
$1/\sqrt{2}(\ket{n}\pm\ket{-n})$ or $1/\sqrt{2}(\ket{n}\pm\ket{-n+1})$ (in
this context $\ket{n}$ denotes the state with particle localized at site $n$)
depending on the type of the mirror-symmetry point. The energy splitting between these two eigenstates can be obtained exactly, and $j_{\rm eff,0}(k)$ is calculated simply as one half of the splitting. The other ingredient needed to evaluate the sum in Eq.~\eqref{S_n_sum} is the integral given by Eq.~\eqref{S_n_int}. In principle, an exact numerical evaluation of this
integral is possible, but for large time $t$ the integrand becomes a highly
oscillatory function of $\phi$ which leads to numerical difficulties. To
circumvent this problem, we sample $2\times10^4$ points for $\phi$, calculate
the time dependence $S_n(t)$ for each of these points analytically based on the
effective model given in Eq.~\eqref{Ham2}, and then average the results. In
Supplementary Fig.~\ref{smfig1}a we show $\overline{S_n}(t)$ obtained after
averaging over $\phi$ and initial states of the whole configurational basis.

To describe the effects of beats, we again average $2\times 10^4$ samples of $S_n(t)$ obtained for $\phi\in[0, 2\pi]$. The dynamics is calculated by diagonalizing the effective Hamiltonian~\eqref{Ham4}, where $\phi$ determines QP field strengths at sites $a$, $b$, $c$, $d$. As in Fig.~\ref{fig1}a, we take pairs of sites $a$, $b$ and $c$, $d$ to be neighbouring, while QP field frequency $\beta$ fixes the distance $d$ between the pairs (e.g.\ $d=2$ for
$\beta=2/(1+\sqrt{5})$). AM appears at a timescale given by $1/\epsilon$ as
discussed in the main text. Importantly, the distribution of values of
$\epsilon$ obtained for different phases $\phi$ is narrow because $\epsilon$ is
set mainly by distance $d$ and only weakly depends on $\phi$. Therefore, we
take $\epsilon$ to be independent of $\phi$ and, to simulate the aforementioned
weak broadening of the distribution, we add $\epsilon$ to the unperturbed
energies $E_i^0$ in Eq.~\eqref{Ham4} instead of adding it to eigenvalues
$E_i$. The value of $\epsilon$ is therefore the only free parameter in our
model and it is chosen so that the same AM timescale as the one obtained for a
single phase $\phi_{\rm PR}$ of point reflection symmetry (see
Figs.~\ref{fig1}a and~\ref{fig2}) is recovered.

Finally, we have to combine single-particle and beat effects in a single model.
We note that the quantum beat model already partly includes single-particle effects, as it includes single-particle hoppings between pairs of sites $a$, $b$, and $c$, $d$. If the boundary between the subsystems is placed between $a$ and $b$, and if $a$ and $b$ are neighbouring as in Fig.~\ref{fig1}a, the quantum beat model reproduces the part of single-particle dynamics that corresponds to nearest-neighbour hopping, as seen from identical short-time dynamics from the single-particle model in Supplementary Fig.~\ref{smfig1}a and from the quantum beat model in Supplementary Fig.~\ref{smfig1}b. To isolate only effects of beats, which occur at much longer timescales, we first average $S_n(t)$ between $t=10^2$ and $t=5\times10^3$, i.e.\ after the single-particle dynamics has saturated and before beats appear, see Supplementary Fig.~\ref{smfig1}b. Then, we add only the $S_n$ growth relative to that average (green line in Supplementary Fig.~\ref{smfig1}b) after $t=5\times10^3$ to the results of the pure single-particle model from Eq.~\eqref{S_n_sum}. For $t<5\times 10^3$ we only take the results of the pure single-particle model. We also note that the choice of the averaging interval is somewhat arbitrary; it is only important that the interval begins after nearest-neighbour hopping dynamics is saturated and before beats become significant. The procedure we outlined gives excellent agreement with exact results obtained from full ED, as shown in Supplementary Fig.~\ref{smfig1}c (identical to Fig.~\ref{fig2}b in the main text).

\end{document}